\documentclass[letter]{aa}
\usepackage[T1]{fontenc}
\usepackage{natbib}
\usepackage{csquotes}
\usepackage{xspace}
\usepackage{txfonts}
\usepackage{amsmath}
\usepackage{threeparttable}
\usepackage{graphicx}
\usepackage{hyperref}
\usepackage[separate-uncertainty=true]{siunitx}
\usepackage{tikz}
\usepackage{subcaption}
\usepackage{placeins}

\newcommand\commit[2]{\node[commit] (#1) {}; \node[clabel] at (#1) {\texttt{#1}: #2};}
\newcommand\connect[2]{\path (#1) to[out=90,in=-90] (#2);}

\def\FINALSIMTIME{\SI{86000}{\day}}
\def\FINALSIMTIMEDRAG{\SI{10000}{\day}}
\usepackage{color}

\begin{document}

\title{No evidence for radius inflation in hot Jupiters from vertical advection of heat}
\titlerunning{No inflation from vertical advection}
\authorrunning{Schneider et al.}
\author{Aaron David~Schneider$^{1,2}$, Ludmila Carone$^{3,4}$, Leen Decin$^{2}$, Uffe Gr\r{a}e~J{\o}rgensen$^{1}$ \& Christiane Helling$^{3}$}
\institute{
  (1) Centre for ExoLife Sciences, Niels Bohr Institute, {\O}ster Voldgade 5, 1350 Copenhagen, Denmark\\
  (2) Institute of Astronomy, KU Leuven, Celestijnenlaan 200D, 3001, Leuven, Belgium\\
  (3) Space Research Institute, Austrian Academy of Sciences, Schmiedlstrasse 6, A-8042 Graz, Austria\\
  (4) Centre for Exoplanet Science, School of Physics \& Astronomy, University of St Andrews, North Haugh, St Andrews, KY169SS, UK\\
} \date{\today}
\offprints{A. Schneider,\\ \email{aaron.schneider@nbi.ku.dk}\\or \email{aarondavid.schneider@kuleuven.be}}
\abstract{Elucidating the radiative-dynamical coupling between the upper photosphere and deeper atmosphere may be key to our understanding of the abnormally large radii of hot Jupiters. Very long integration times of 3D general circulation models (GCMs) with self-consistent radiative transfer are needed to obtain a more comprehensive picture of the feedback processes between dynamics and radiation. Here, we present the longest 3D nongray GCM study to date (\FINALSIMTIME) of an ultra-hot Jupiter (WASP-76 b) that has reached a final converged state. Furthermore, we present a method that can be used to accelerate the path toward temperature convergence in the deep atmospheric layers. We find that the final converged temperature profile is cold in the deep atmospheric layers, lacking any sign of vertical transport of potential temperature by large-scale atmospheric motions. We therefore conclude that coupling between radiation and dynamics alone is not sufficient to explain the abnormally large radii of inflated hot gas giants.}


\maketitle

\section{Introduction}\label{sec:introduction}
Characterizing and understanding the atmospheres of exoplanets is now more within reach than ever before thanks to the recent launch of the \textit{James Webb Space Telescope} (JWST). Many of the observed hot gas giants exhibit very large radii and therefore low bulk densities. One of the long-standing questions in the field of hot and ultra-hot Jupiters (UHJs) pertains to the mechanism that inflates those planets. Interior models of the observed exoplanet population reveal that the state of inflation links the incident flux with the temperature in the deep atmosphere \citep{Thorngren2018Inflation,Thorngren2019Interior,Sarkis2021Interior}. These studies find that the intrinsic temperatures increase with instellation and peak at an instellation of $\approx\SI{2e9}{erg\per\s\per\cm\squared}$ \citep{Thorngren2019Interior}. Different explanations have been proposed to understand the mechanisms that lead to these hot intrinsic temperatures. There are two hypotheses that could explain inflation \citep{Fortney2021review}. The first proposes that planets are heated by tidal interactions \citep[e.g.,][]{Bodenheimer2001,ArrasSocrates2010Tides,Socrates2013Tides}, whereas the second claims that incident stellar flux is deposited into the deep atmosphere.

A prominent theory that fits with the latter hypothesis is Ohmic dissipation \citep{Perna2010bOhm,Batygin2010Inflation,Menou2012Ohm,Rauscher2013Ohm,Helling2021CloudsUHJ, Knierim2022Interior}, which proposes that magnetic fields should couple to ionized flows in the upper atmosphere, slowing down zonal winds and therefore depositing heat by friction. However, it is unclear whether or not this mechanism is sufficiently active at sufficiently deep layers to heat the deep atmosphere \citep{Rauscher2013Ohm}. Using a 2D model with parameterized vertical transport, \citet{Tremblin2017Inflation} find that heat would be transported downwards by means of vertical transport of potential temperature without the need of magnetic fields. The authors claim that incident energy input from stellar instellation is self-consistently advected downwards. Idealized 3D general circulation models (GCM) with parametrized thermal forcing seem to reproduce this mechanism \citep{Sainsbury-Martinez2019Inflation,Sainsbury-Martinez2021Inflation}. 

In recent work, we introduced \texttt{expeRT/MITgcm,} a 3D GCM with self-consistent radiative transfer for hot Jupiters without the limitations of the above-mentioned models \citep{Schneider2022GCM}. In this latter paper, we focussed on the effects of surface gravity and rotation rate on the temperature evolution. Here, we extend this work to the final state of the converged atmosphere, using the same 3D nongray GCM and integrating for long enough to converge the temperature  in the deep atmosphere. 

In order to find a planet that is suitable for studying the issue of temperature convergence in the deep layers of the atmosphere, we decided to go to the temperature extremes and simulate the ultra-hot Jupiter WASP-76 b. WASP-76 b has been observed with radial velocity \citep{West2016W76Obs} and transit spectroscopy \citep{Seidel2019W76b} and has both a Spitzer phase curve \citep{May2021GCM} and winds characterized by high-resolution spectroscopy \citep{Ehrenreich2020W76,Kesseli2021W76,Kesseli2022W76}. Additionally, the climate of WASP-76 b was recently modeled \citep{Wardenier2021W76,Savel2022GCM,Beltz2022GCMObs}. WASP-76 b has a very low bulk density and can therefore be considered to be inflated, while we also know that it receives a high incident stellar flux. We therefore expect there to be a mechanism at play that inflates the radius of this giant planet by transporting potential energy downwards.

We begin by describing the methods used in this paper (Sect.~\ref{sec:methods}). We then explain how we reached temperature convergence in our models (Sect.~\ref{sec:results}) before discussing the limitations of our methods (Sect.~\ref{sec:discussion}). Finally, we present our conclusions in Sect.~\ref{sec:conclusion}.


\section{Methods}\label{sec:methods}
\subsection{General circulation model}
This work utilizes \texttt{expeRT/MITgcm} described in detail in \citet{Schneider2022GCM} and \citet{Carone2020GCM}. In particular, \texttt{expeRT/MITgcm} builds on the dynamical core of the general circulation model (GCM) \texttt{MITgcm} \citep{Adcroft2004} and solves the 3D hydrostatic primitive equations of fluid dynamics \citep[e.g.,][Eq.~1-4]{Showman2009GCM} on an Arakawa C-type cubed-sphere grid with a resolution of C32 assuming an ideal gas. The vertical grid contains 41 logarithmically spaced layers from \SI{1e-5}{\bar} to \SI{100}{\bar}, extended by six linearly spaced layers between \SI{100}{\bar} and \SI{700}{\bar}. The model includes Rayleigh friction at the bottom of the computational domain (below \SI{490}{\bar}), a sponge layer at the top of the atmosphere (above \SI{1e-4}{\bar}), and a fourth-order Shapiro filter.  We caution that the choice of numerical methods, such as the choice of the dynamical core, often has a significant influence on the circulation and the temperature \citep{Heng2011Benchmark,Polichtchouk2014GCM,Skinner2021GCM,Carone2021}. For the scope of this work, we identify the use of deep drag as having possible repercussions for our findings, and therefore discuss the influence of the drag on the results of this work in Appendix~\ref{sec:drag}.

The temperature in the atmosphere is forced by radiative heating and cooling using a multi-wavelength radiative transfer scheme that operates during the runtime of the climate model. More specifically, the radiation field is updated with a radiative time step of $\Delta t_\mathrm{rad}=\SI{100}{\s}$, which is four times the dynamical time step $\Delta t_\mathrm{dyn}=\SI{25}{\s}$. The radiative transfer includes scattering and uses five correlated-k wavelength bins with 16 Gaussian quadrature points per wavelength bin. The implementation of radiative transfer in \texttt{expeRT/MITgcm} is based on \texttt{petitRADTRANS} \citep{Molliere20191Dmodel, Molliere20201Dmodel}. One advantage of using  \texttt{expeRT/MITgcm} is its flexibility and performance, which enable long integration times while maintaining the accuracy of a multi-wavelength radiation scheme \citep{Schneider2022GCM}. 

We use opacities on a pre-calculated grid of pressure and temperature, assuming local chemical equilibrium. Most opacities are obtained from exomol\footnote{\url{https://www.exomol.com/}}, and we use H$_2$O \citep{Polyansky2018Opac},  CO$_2$ \citep{Yurchenko2020Opac}, CH$_4$ \citep{Yurchenko2017Opac}, NH$_3$ \citep{Coles2019Opac}, CO \citep{Li2015Opac}, H$_2$S \citep{Azzam2016Opac}, HCN \citep{Barber2014Opac}, PH$_3$ \citep{Sousa-Silva2015Opac}, TiO \citep{McKemmish2019Opac}, VO \citep{McKemmish2016Opac}, FeH \citep{Wende2010Opac}, Na \citep{Piskunov1995Opac}, and K \citep{Piskunov1995Opac} opacities for the gas absorbers. Furthermore, we include Rayleigh scattering with H$_2$ \citep{Dalgarno1962Opac} and He \citep{Chan1965Opac} and collision-induced absorption (CIA) with H$_2$--H$_2$ and H$_2$--He \citep{Borysow1988Opac, Borysow1989aOpac, Borysow1989bOpac, Borysow2001Opac, Richard2012Opac, Borysow2002Opac} and H$^-$ \citep{Gray2008Opac}.

As the goal of this paper is to look at the conservation of energy and the interplay between dynamical energy transport and radiative energy transport, we updated \texttt{expeRT/MITgcm} to conserve energy lost by friction at the bottom boundary. Kinetic energy lost due to friction on the horizontal velocity $\vec u$ [\SI{}{\m\per\s}] is converted to thermal energy using the following equation \citep{Rauscher2013Ohm}:
\begin{equation}
        \frac{\mathrm{d}T}{\mathrm{d}t} = \frac{u^2}{c_p \tau_\mathrm{deep}},
\end{equation}
where $\frac{\mathrm{d}T}{\mathrm{d}t}$ [\SI{}{\K\per\s}] is the change in the temperature $T$ [\SI{}{\K}], $\tau_\mathrm{deep}$ is the friction timescale \citep[as described in][]{Schneider2022GCM, Carone2020GCM}, and $c_p$ [\SI{}{\J\per\kg\per\K}] is the heat capacity at constant pressure. Otherwise, we use the same parameters as in \citet{Schneider2022GCM}. 

\subsection{Setup of WASP-76 b simulations}
\begin{figure}
\centering
\begin{tikzpicture}
\tikzstyle{commit}=[draw,circle,fill=white,inner sep=0pt,minimum size=5pt]
\tikzstyle{clabel}=[right,outer sep=1em]
\tikzstyle{every path}=[draw]
\matrix [column sep={1em,between origins},row sep=\lineskip]
{
\commit{0d}{start} & \\
\commit{40000d}{split simulations} & \\
 & \commit{steroids}{starting off from previous prediction} \\
\commit{nominal}{continuing integration} \\
};
\connect{40000d}{0d}
\connect{nominal}{40000d}
\connect{steroids}{40000d}
\end{tikzpicture}
\caption{Overview of the simulations in this work. The \texttt{steroids} model starts at $t=\SI{40000}{\day}$ from the \texttt{nominal} model with the guess of the final state obtained via the method outlined in Appendix \ref{sec:ca}.}
\label{fig:vis}
\end{figure}
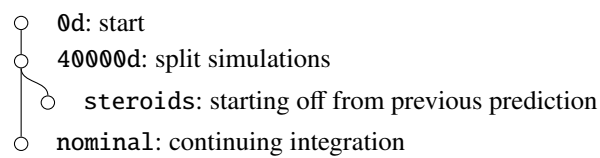
We use \texttt{expeRT/MITgcm} to model WASP-76 b, an ultra-hot Jupiter orbiting WASP-76, where we assume $T_\star=\SI{6250}{\K}$ and $R_\star=\SI{1.73}{R_\odot}$ as the stellar temperature and radius of WASP-76  \citep{West2016W76Obs}. We model WASP-76 b on a tidally locked orbit of \SI{0.033}{AU} with an orbital period of \SI{1.8}{\day}  \citep{West2016W76Obs}. The planetary radius and mass are taken as $1.83 R_\mathrm{jup}$ and $0.92 M_\mathrm{jup}$ respectively  \citep{West2016W76Obs}, resulting in a low gravity of $g=\SI{6.81}{\m\per\s\squared}$. We use $c_p=\SI{13784}{\J\per\kg\per\K}$ and $R=\SI{3707}{\J\per\kg\per\K}$ as the specific heat capacity at constant pressure and the specific gas constant, respectively.

We set up two models in this paper (Fig.~\ref{fig:vis}). The \texttt{nominal} model is a model of WASP-76 b, which has been initialized with a particularly hot initial state. We use the same method to initialize the temperature profile that was used in \citet{Schneider2022GCM}, incorporating an adiabatic temperature profile below $\SI{10}{\bar}$ with $\Theta_\mathrm{ad}=\SI{4000}{\K}$ as the temperature that the adiabat would have at \SI{1}{\bar}. We chose such a hot initial state because planets are thought to form in hot conditions \citep[e.g.,][]{Fortney2021review}.

The \texttt{steroids} model has been initialized from the \texttt{nominal} model at $t=\SI{40000}{\day}$. The initialization of the \texttt{steroids} model is done by extrapolating the temperature evolution of the \texttt{nominal} model using log-linear regression. The extrapolation to find the \texttt{steroids} model resembles a hotter limit on the prediction from the temperature evolution of the \texttt{nominal} model up until $t=\SI{40000}{\day}$, while we take over the dynamical state (e.g., the velocity field). We explain the details of the fit and initialization in Appendix~\ref{sec:ca}.

\section{Results from nongray 3D GCM}\label{sec:results}

\begin{figure}
        \includegraphics{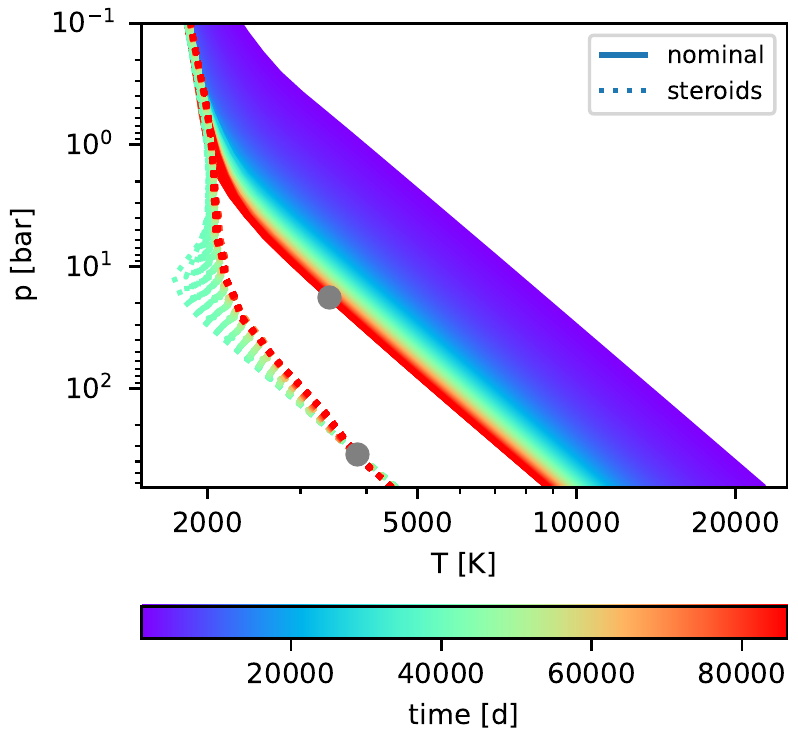}
        \centering
        \caption{Temperature evolution (color coded) of the horizontally averaged temperature in the deep layers of the atmosphere for the \texttt{nominal} model (solid lines) and the \texttt{steroids} model (dotted lines). A gray dot displays the position of the RCB at the final state.}
        \label{fig:timeevol}
\end{figure}
\begin{figure}
        \includegraphics[width=.45\textwidth]{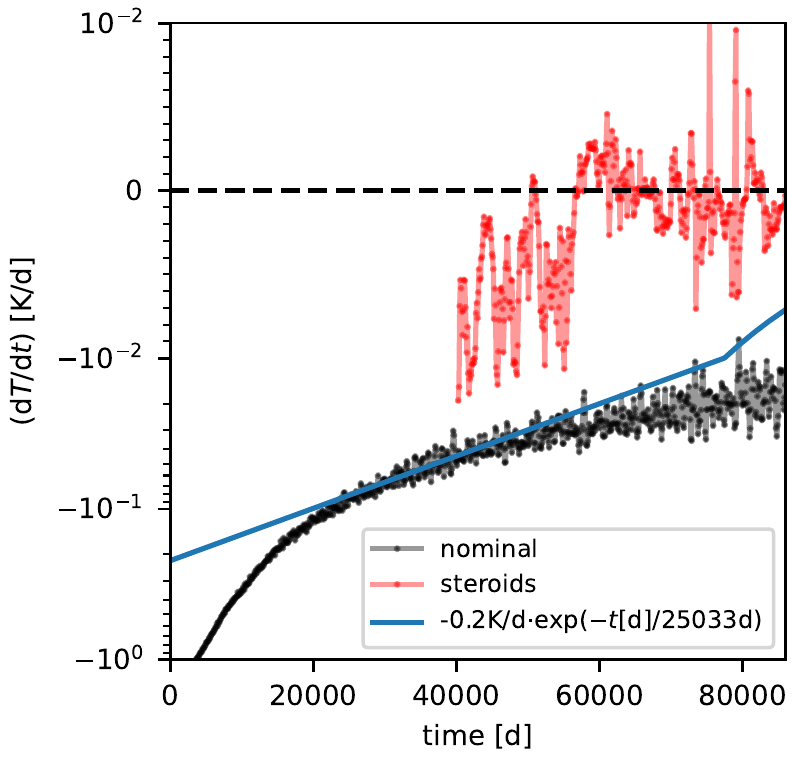}
        \centering
        \caption{Evolution of the rate of temperature change at \SI{650}{\bar} with time of the \texttt{nominal} (black) and the  \texttt{steroids} (red) model. We note that the y-axis is linear between \num{-0.01} and \num{0.01} and logarithmic elsewhere. The fit on the rate of temperature change by an exponential decay around $t=\SI{40000}{\day}$ is shown in blue.}
        \label{fig:timeevol_650}
\end{figure}

Utilizing the performance of \texttt{expeRT/MITgcm}, we monitor the temperature convergence of WASP-76 b up to $t=\FINALSIMTIME$. We show the final atmospheric state and observables in Appendix~\ref{sec:atmo}. The horizontally averaged thermal profile for pressures above $p=\SI{0.1}{\bar}$ is shown in Fig.~\ref{fig:timeevol}. The \texttt{nominal} model cools down from an initially hot state to an almost stable solution within \SI{30000}{\day}, where the temperature changes at the bottom boundary become smaller than $\SI{0.1}{\K\per\day}$, as shown in Fig.~\ref{fig:timeevol_650}. However, the temperature of the model continues to decrease further and will reach a much colder state within reasonable timescales. Structure models such as those of \citet{Thorngren2019Interior} and \citet{Sarkis2021Interior} often give predictions about the location of the radiative convective boundary (RCB), defined as the first pressure layer from the bottom upwards, where the temperature becomes subadiabatic \citep{Thorngren2019Interior}. For better comparison of our model to those predictions, we calculate the RCB in the same way, and show this as gray dots in Fig.~\ref{fig:timeevol}.

We fitted the temperature evolution around $\SI{40000}{\day}$ in order to estimate how much colder the final state would be. Assuming an exponential decay of the temperature changes with time, we estimated the final temperature at $p=\SI{650}{\bar}$ to be $T=\SI{4567}{\K}$ (Appendix~\ref{sec:ca}). We performed a second simulation (dubbed \texttt{steroids}) using this final state as an initial condition to monitor the stability of such a model. The resulting temperature evolution of this model is shown as dotted lines in Fig.~\ref{fig:timeevol}. The new initial state seems to be a good initial guess in the deep layers, whereas some adjustment happens in intermediate layers between \SI{0.1}{\bar} and \SI{100}{\bar}. Looking at the temperature change rates (Fig.~\ref{fig:timeevol_650}), we find that the temperature reaches a steady state, where the temperature stops changing, and the temperature changes start to swap signs in an oscillating manner with \SI{\approx 5e-3}{\K\per\day}. We therefore claim that temperature convergence has been reached in the \texttt{steroids} model, implying that the temperature of our WASP-76 b models self-consistently converges to a cold interior. 

The planet gains energy from stellar irradiation and loses energy by thermal radiation. We can therefore formulate the energy balance as an integral over the net radiative flux $F_\mathrm{tot}$ as
\begin{equation}\label{eq:energy_balance}
        \int F_\mathrm{tot}(\varphi,\vartheta)\mathrm{d}A = \int (F_\star+F_\mathrm{pla})\mathrm{d}A  = 4\pi R_\mathrm{pla}^2\sigma_\mathrm{SB}T_\mathrm{int}^4,
\end{equation}
where $\vartheta$ and $\varphi$ are latitude and longitude respectively, $T_\mathrm{int}$[K] is the intrinsic temperature and $R_\mathrm{pla}$ is the planetary radius. Here, $F_\mathrm{pla}$ and $F_\star$ are the planetary and stellar fluxes, respectively. In radiative equilibrium, without an internal energy source, it should hold that 
\begin{equation}
        \int F_\star\mathrm{d}A = -\int F_\mathrm{pla}\mathrm{d}A,
\end{equation}
and subsequently $T_\mathrm{int} = \SI{0}{\K}$ from Eq.~\ref{eq:energy_balance}. We note that we do not impose a nonzero flux as the boundary condition of the radiative transfer in our GCM. Instead, the boundary condition for the thermal flux at the deepest layer of the atmosphere is zero, which corresponds to $T_\mathrm{int} = \SI{0}{\K}$.

\begin{figure*}
                \centering
                \includegraphics{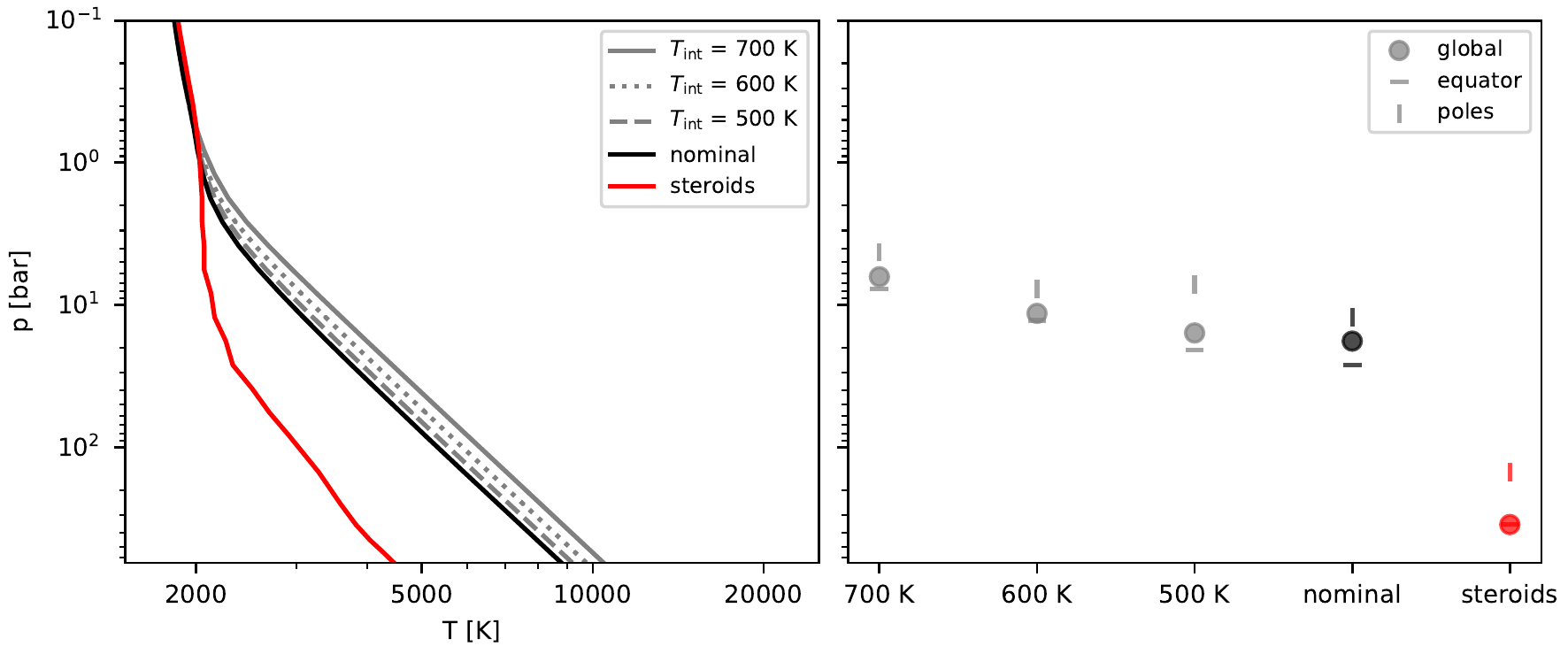}
                \caption{Temperature profiles and RCB locations. Left: Temperature profiles for different intrinsic temperatures (gray) and final temperature profiles of the \texttt{nominal} ($T_\mathrm{int}\approx\SI{434}{\K}$) and \texttt{steroids} simulation. Right: Calculated RCB for the different temperature profiles of the left panel. Additional markers display the RCB calculated for the equatorial and the polar region. The gray profiles are calculated from the temperature evolution of the \texttt{nominal} model. We note that the \texttt{steroids} model loses energy due to dissipation and therefore has no physical intrinsic temperature. 
                }
                \label{fig:globT_with_Tint}     
\end{figure*}

We display the effect of the intrinsic temperature on the temperature profile in Fig.~\ref{fig:globT_with_Tint}, where we show temperature profiles with three different intrinsic temperatures (\SI{500}{\K}, \SI{600}{\K,} and \SI{700}{\K}) along with the final time steps of the \texttt{steroids} and \texttt{nominal} model from Fig.~\ref{fig:timeevol}. The temperature profiles for the different intrinsic temperatures have been calculated from the temperature evolution of the \texttt{nominal} model (Fig.~\ref{fig:timeevol}) using Eq.~\ref{eq:energy_balance} with bolometric fluxes, which was computed during the runtime of the climate simulations. The \texttt{steroids} model has a negative total net flux, rendering it impossible to calculate a physical meaningful intrinsic temperature.

Combining the predictions from \citet{Thorngren2019Interior} and \citet{Sarkis2021Interior}, we find that the intrinsic temperature of WASP-76 b should be approximately $\SI{600\pm100}{\K}$ and the RCB location should be \SI{1\pm1}{\bar} according to structure models. Thus, we can compare the intrinsic temperature to the predictions and find that the intrinsic temperature of the final \texttt{nominal} model is $\SI{434}{\K}$, which is significantly below the predicted value of $\SI{600}{\K}$ from structure models. 

The right plot in Fig.~\ref{fig:globT_with_Tint} displays the RCB location for the same models. Comparing the RCB location of the final globally averaged \texttt{nominal} and \texttt{steroids} model to the predictions from the above-mentioned 1D structure models, we find that the locations of the RCB in our models are much deeper than 10 bar, and therefore not in line with predictions. \citet{RauscherShowman2014GCM} showed that the RCB location varies horizontally, because the equatorial regions have higher stellar fluxes than the polar regions. We therefore performed additional calculations of the RCB with temperature averages of the polar ($|\vartheta|>45^\circ$) and equatorial ($|\vartheta|<45^\circ$) regions. We can confirm the findings of \citet{RauscherShowman2014GCM}, namely that the RCB location seems to vary horizontally, where we find higher and deeper levels for polar and equatorial regions, respectively. The different values of the RCB for different horizontal regions is caused by the fact that the temperature becomes horizontally homogeneous only at deeper levels in the atmosphere. We therefore conclude that the intrinsic model differences of 1D and 3D models make the comparison of RCB locations more difficult.

\begin{figure*}
        \includegraphics{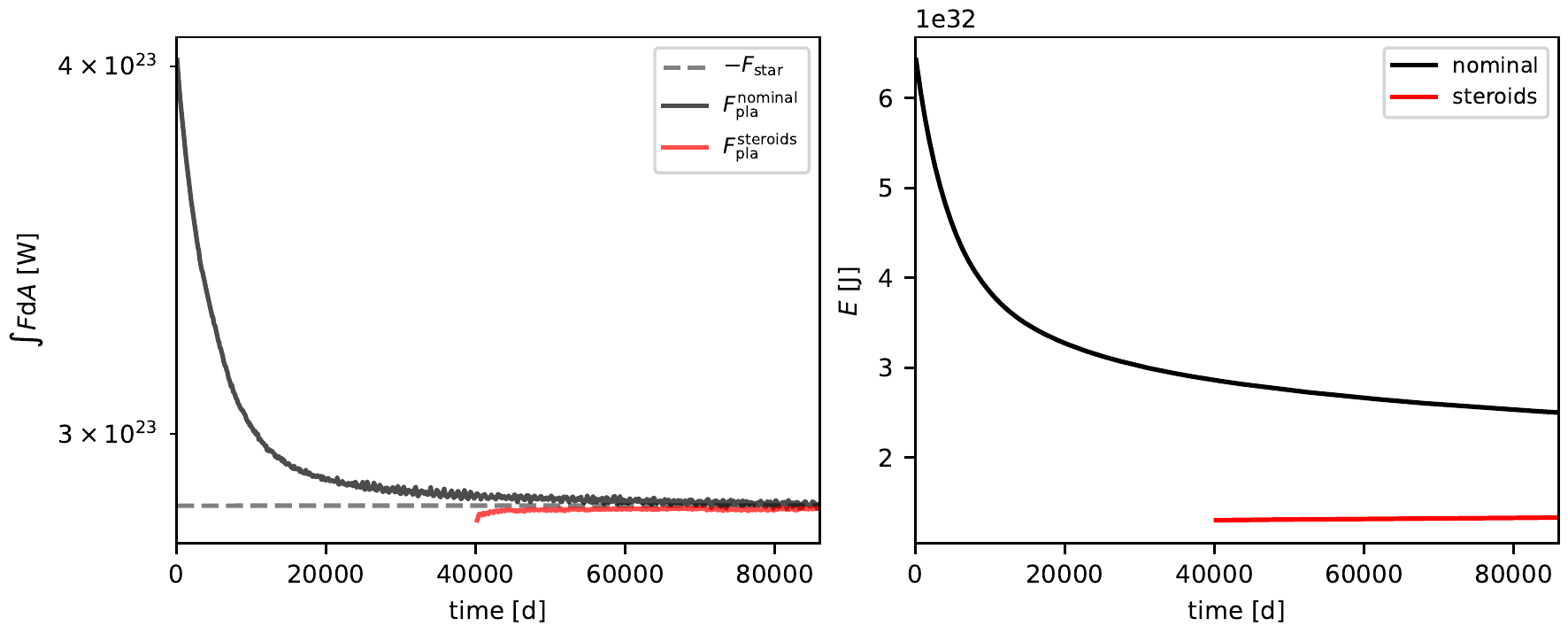}
        \centering
        \caption{Energy budget of the \texttt{nominal} and \texttt{steroids} simulations. Left: Radiative energy budget $\left(\int F \mathrm{d}A\right)$. Right: Evolution of the total energy. Both simulations converge toward radiative equilibrium with approximately 0.1\% accuracy. }
        \label{fig:flux_diff}
\end{figure*}
We show the evolution of the area-integrated radiative fluxes ($F_\star$ and $F_\mathrm{pla}$) along with the total energy of the \texttt{steroids} and \texttt{nominal} in Fig.~\ref{fig:flux_diff}. 
We follow \citet{Polichtchouk2014GCM} and calculate the total energy (TE) as a sum of the potential, kinetic, and internal energy by
\begin{equation}
        \mathrm{TE} = \int_V \left(\frac{u^2+v^2}{2}+c_pT+gZ\right) \rho\mathrm{d}V,
\end{equation}
where $Z$ is the geometric height.
The plot shows that the high temperatures in the deep atmosphere do not lead to large flux differences between the stellar and planetary fluxes, which can be explained by the relation between flux and temperature, which goes with the fourth power of the temperature (Eq.~\ref{eq:energy_balance}). Intrinsic temperatures on the order of \SI{1000}{\K} therefore become insignificant against the temperature of the thermal emission, which is on the order of \SI{2000}{\K}. The \texttt{nominal} model loses energy during its evolution as the planet cools down, whereas the \texttt{steroids} model heats up again  slightly. These trends manifest themselves in the right panel of Fig.~\ref{fig:flux_diff}, showing that the \texttt{steroids} model conserves the total energy, whereas the \texttt{nominal} model is still losing energy through radiation to space.

Our model conserves the radiative energy to roughly 99.9\%. \citet{Deitrick2022GCM} recently reported that in practice their \texttt{THOR} GCM always tends toward a negative radiative flux balance, with discrepancies of as large as a few percent. The authors claim that the reason for that can be found in the numerical dissipation processes in their model such as sponge layers and hyper-diffusion.\footnote{We note that all GCMs need some form of dissipation for numerical reasons \citep{Jablonowski2011}.} Similarly, \citet{Rauscher2012double} report that their total radiative heating rate is always positive, balancing a loss of kinetic energy due to numerics. The conservation of energy and angular momentum in \texttt{MITgcm} was benchmarked by \citet{Polichtchouk2014GCM}, revealing that \texttt{MITgcm} is particularly poor at conserving angular momentum, whereas it performs well at conserving energy, which can be confirmed for the \texttt{steroids} model in the right panel of Fig.~\ref{fig:flux_diff}. We therefore think that conservation of the radiative energy with an accuracy of 99.9\% is already very good. It is therefore perfectly reasonable that the \texttt{steroids} model is indeed losing more energy than it is gaining, while still being in equilibrium. 

\section{Discussion}\label{sec:discussion}

We conclude that we cannot find large-scale atmospheric motions able to inflate a planet significantly. Instead, we find that heating and cooling by radiation dominate the temperature evolution even in deep layers, forcing our nongray simulations toward a cool final state. We therefore propose that hydrodynamical models of hot gas giants need extra physics or parameterization (such as dissipative drag or diffusion) to recreate an inflated interior. 

\citet{YoudinMitchel2010} and \citet{Tremblin2017Inflation} showed that vertical heat transport by turbulent mixing could possibly inflate the interior. However, their models do not model vertical mixing self-consistently, instead relying on a parameter that sets the mixing strength. Using idealized GCM simulations, \citet{Sainsbury-Martinez2019Inflation} claimed that large-scale atmospheric motions could be a reason for a tendency toward a hot adiabatic interior in their simulations. There are two major differences between the model used in this work and the GCM used in the work of these latter authors that need to be considered when questioning the lack of a hot interior in our models in comparison to their work. One of the obvious differences is the treatment of radiation: whereas  \citet{Sainsbury-Martinez2019Inflation} do not include radiative cooling and heating in the deep atmosphere, we apply a nongray radiative transfer scheme. 

Perhaps equally as important, the second difference is the use of numerical dissipation, where  \citet{Sainsbury-Martinez2019Inflation} use a hyper-diffusion scheme to stabilize their model, whereas we use a fourth-order Shapiro filter. Previous studies showed that in general, both dissipation schemes (filtering and diffusion) have a profound influence on the temperature and dynamics of a GCM \citep[e.g.,][]{Polichtchouk2014GCM,Koll2018GCMDiffusion,Hammond2022GCMDiffusion}. \citet{Sainsbury-Martinez2019Inflation} showed that the strength of numerical diffusion in their model critically impacts the timescale of the tendency toward the hot adiabat, where a strong numerical diffusion relates to a faster evolution toward a hot interior. It could therefore be possible that their finding might have been caused by heat diffusion from their numerical dissipation scheme.

It is still unclear how the interior of hot and ultra-hot Jupiters interact with the upper atmosphere. There are currently few studies that link the convective interior with the photosphere dynamics \citep{Showman2020review}. The first steps toward this goal were made by \citet{Lian2022InteriorGCM}, who model the effect of convective forcing on the circulation patterns of hot irradiated planets with a combination of Newtonian forcing in the upper atmosphere and perturbations in the deep atmosphere. A model able to combine 1D interior models with realistic 3D atmospheric models including radiative transfer would also be desirable and would reveal insights into the connection between the convection in the deep interior and the atmosphere. Similar approaches, connecting convective interiors with upper radiative atmospheres, exist in the stellar community \citep[e.g.,][]{Skartlien2000waves, Freytag2010Convection,Freytag2017ConvectionAGB,Freytag20193DAGB}.

The present work is the most complete GCM study to date, which is aimed at understanding the coupling of the upper and the deeper atmosphere in hot Jupiters. However, we note that we are so far omitting important effects for ultra-hot Jupiters: Ohmic dissipation, dissociation of H$_2$ and clouds. The upper atmospheres of ultra-hot Jupiters like WASP-76b are highly ionized due to their high temperatures \citep[e.g.,][]{Helling2021CloudsUHJ}. Magnetic fields, if present, would therefore couple strongly to the ionized winds. \citet{Rauscher2013Ohm} and \citet{Beltz2022GCMObs} showed that friction induced by magnetic fields acting on ionized winds may alter the dynamical state of the atmosphere and lead to additional heating in the upper atmosphere, where friction is converted to heat. It is still debated and unclear \citep{Rauscher2013Ohm,Showman2020review} as to whether this extra heat could be sufficient to inflate a planet. 

Recent simulations of ultra-hot Jupiters showed that H$_2$ dissociation can alter their thermal structure \citep{Tan2019drag}. The energy needed to dissociate molecular hydrogen on the day side cools the atmosphere, while recombination of atomic hydrogen heats the night side, leading to an overall reduced day--night temperature contrast. Similarly, the presence of clouds in a 3D GCM will alter the thermal and dynamical state of hot Jupiters \citep{Roman2021Clouds,Deitrick2022GCM}. Realistic models of ultra-hot Jupiters should therefore take these effects into account, even though it is unlikely that these effects are important for the deeper parts of the atmosphere.

\section{Summary and conclusions}\label{sec:conclusion}

In this work, we investigate one of the main hypotheses proposed by \citet{Guillot2002Interior} and \citet{YoudinMitchel2010} to explain the abnormally large radii of inflated hot Jupiters: Energy transport from the irradiated atmosphere into deeper layers. We performed long-term (\FINALSIMTIME) nongray 3D GCM simulations of WASP-76 b to search for signs of vertical advection of potential temperature from the upper irradiated atmosphere into the interior. Our simulations started from a hot initial state and cooled down toward a colder state. We then used an exponential fit on the temperature convergence to extrapolate the temperature toward a possible final converged state, which was reached soon after for the extrapolated simulation. The final converged simulation exhibits a cold temperature profile, lacking signs of vertical advection of heat from the upper atmosphere into the interior. We find instead that the atmosphere has cooled down to radiative equilibrium, conserving energy with an accuracy of 99.9\%. 

Our work strongly disfavors vertical downward advection of energy from the irradiated part of the atmosphere by large-scale atmospheric circulation as a possible explanation for inflation. We suggest instead that physical processes other than radiation and dynamics need to be taken into account in order to match the abnormal large radii of inflated hot Jupiters. Such processes could be the interaction between a convective interior and the atmosphere, small-scale turbulent transport, or magnetic field interactions. Future investigations are needed to quantify their effects on the state of radius inflation.

\begin{acknowledgements}
We thank Xianyu Tan for fruitful discussions on the effect of numerical dissipation on the thermal structure of GCM simulations. We also thank an anonymous referee for their useful comments that improved the quality of the manuscript.
A.D.S., L.D., U.G.J and C.H. acknowledge funding from the European Union H2020-MSCA-ITN-2019 under Grant no. 860470 (CHAMELEON). L.C. acknowledges the Royal Society University Fellowship URF R1 211718 hosted by the University of St Andrews. U.G.J acknowledges funding from the Novo Nordisk Foundation Interdisciplinary Synergy Program grant no. NNF19OC0057374. The bibliography of this publication has been typesetted using \texttt{bibmanager} \citep{bibmanager}\footnote{\url{https://bibmanager.readthedocs.io/en/latest/}}. The post-processing of GCM data has been performed with \texttt{gcm-toolkit} \citep{gcm_toolkit} \footnote{\url{https://gcm-toolkit.readthedocs.io/en/latest/}}
\end{acknowledgements}

\bibliographystyle{aa}
\bibliography{ms}

\begin{appendix}
\section{Effect of the drag}\label{sec:drag}
\begin{figure}
                \centering
                \includegraphics{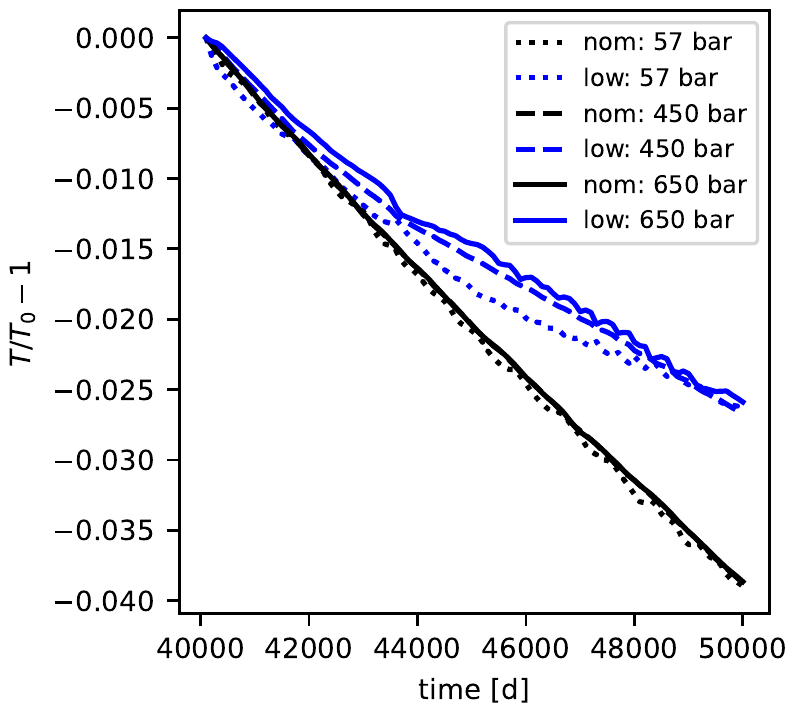}
                \caption{Evolution of the temperature in three pressure layers (\SI{57}{\bar}, \SI{450}{\bar}, \SI{650}{\bar}) in a simulation with low drag compared to the \texttt{nominal} simulation. The temperature in each layer is normed to the respective initial value at \SI{40000}{\day}.}
                \label{fig:ld_T}        
\end{figure} 
\begin{figure}
                \centering
                \includegraphics{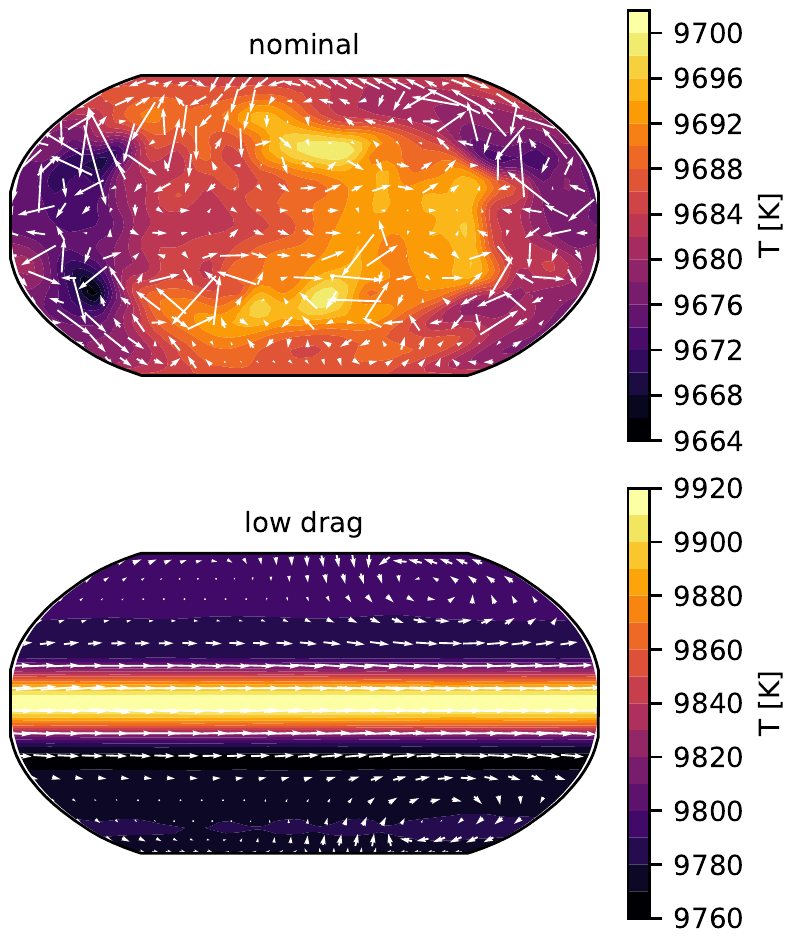}
                \caption{Isobaric temperature slices at $t=\SI{50000}{\day}$ at $p=\SI{650}{\bar}$ for the \texttt{nominal} and the model with low drag. The temperature has been averaged over the last \SI{100}{\day}. The substellar point is located in the center of the plot.}
                \label{fig:ld_T_drag}   
\end{figure} 
Previous studies revealed the enormous impact of the chosen (often unjustified) dissipation schemes \citep{Heng2011Benchmark, Skinner2021GCM}. We perform a test simulation with a 1000 times longer friction timescale ($\tau_\mathrm{deep}$) compared to the \texttt{nominal} and \texttt{steroids} simulations. This longer friction timescale weakens the strength of the Rayleigh drag applied to the deep atmosphere and makes sure that the prescription of the bottom drag does not affect the conclusions of this work. This additional simulation has been started from the \texttt{nominal} model at \SI{40000}{\day} and has been carried out for \FINALSIMTIMEDRAG.

We compare the temperature evolution of the \texttt{nominal} simulation to the simulation with the low drag in Fig.~\ref{fig:ld_T}, where we find that the  simulation with the lower drag cools down slower than the \texttt{nominal} simulation.  The reason for the difference in the temperature evolution can be found in the Fig.~\ref{fig:ld_T_drag}, where we show the temperature maps at \SI{650}{\bar} of the model with low drag and the \texttt{nominal} model. Both models start from an average temperature of \SI{10080}{\K} at \SI{40000}{\day} and cool down by a few hundred Kelvin. However, we find that the drag stops the equatorial jet, which otherwise descends toward the bottom boundary in the simulation with low drag. Simultaneously, stopping the equatorial jet leads to an efficient north--south redistribution of the temperature at the bottom boundary. Thus, in the model with much smaller drag, we find that the temperature starts diverging between the poles and the equator, where the equator cools down slower than the poles.

Considering this result, it is reasonable to conclude that the final converged solution of a model with lower drag would be hotter than the \texttt{steroids} model, possibly reducing the gap between outgoing and incoming radiation in the left panel of Fig.~\ref{fig:flux_diff}. Nevertheless, we also find that this simulation continues to cool down,  still showing no sign of the vertical advection of heat that would, on the contrary, lead to a warming of the deep layers.

\FloatBarrier
\section{Prediction of the final state of the atmosphere}\label{sec:ca}

Temperature convergence in the deep layers of the atmospheres of hot Jupiters  is impossible to reach in reasonable model runtimes. We therefore propose to speed up convergence by predicting the final converged state and extrapolating to this final state, which can then be used to initialize a new model. The method outlined below is one possible approach to doing this. In order to be able to predict the final state of the temperature in the deep atmosphere, we apply three assumptions:

\begin{enumerate}
        \item The temperature in the deep atmosphere is horizontally isotropic and homogeneous.
        \item The temperature decrease can be modeled by an exponential decay.
        \item The deep layers are adiabatic: $T=T_{\SI{650}{\bar}} \left(\frac{p}{\SI{650}{\bar}}\right)^{R/c_p}$.
\end{enumerate}

We note that these assumptions (especially the second) are rather too simplified to capture the actual cooling of planets, which is not thought to be consistent over time \citep{GinzburgSari2015Interior}. However, these simplifications do not need to be true, because their only purpose is to predict the initial state of the \texttt{steroids} model. Moreover, in order to minimize temporal variations unrelated to the general trend of temperature convergence, we compute the horizontally and temporally averaged temperature over a sampling period of \SI{1000}{\day}. We perform this computation for four sampling periods (e.g., for a total of \SI{4000}{\day}) yielding four values for the potential temperature of four consecutive sampling periods. This computation is done in retrograde at $t=\SI{40000}{\day}$ with the \SI{100}{\day} time-averaged temperature fields from \SI{36000}{\day} to \SI{40000}{\day}.

We then use the least-squares method to fit a linear relation
\begin{equation}
        y_i = \log\left(\frac{(\mathrm{d}T/\mathrm{d}t)_i}{\lambda}\right),
\end{equation}
where $\lambda$ is a norm that ensures that the value inside log stays positive and unitless. We set
\begin{equation}
        \lambda = \sum_j{(\mathrm{d}T/\mathrm{d}t)_j},
\end{equation}
the subscripts $i,j\in\{1,2,3,4\}$ to denote the four support points of the fit. We can now use the least-squares method to solve
\begin{equation}
        \begin{pmatrix}
                y_1\\y_2\\y_3\\y_4
        \end{pmatrix} =
        \begin{pmatrix}
                t_1 & 1\\
                t_2 & 1\\
                t_3 & 1\\
                t_4 & 1
        \end{pmatrix}
        \cdot
        \begin{pmatrix}
                m & c
        \end{pmatrix},
\end{equation}
in order to obtain the best-fit values for $m$ and $c$, which is equivalent to
\begin{equation}\label{eq:fit}
        \frac{\mathrm{d}T}{\mathrm{d}t} = \lambda \tilde c\cdot  \exp(mt),
\end{equation}
where $\tilde c = \exp(c)$. The fit, as displayed in Fig.~\ref{fig:timeevol_650}, matches the temperature change rates around $t=\SI{40000}{\day}$. However, the real temperature change rates for later times are slightly below the fit, meaning that the temperature cools down even faster than predicted by the fit.

We then estimate the final temperature from integrating Eq.~\ref{eq:fit} over time and get
\begin{equation}
        T(t) = T_0 + \frac{\lambda\tilde c}{m}(\exp(mt)-1).
\end{equation}
We evaluate this equation at the age of the mother star, which is $\approx \SI{5}{Gyr}$ \citep{West2016W76Obs}, and find that the temperature at \SI{650}{\bar} should be $T_{\SI{650}{\bar}} = \SI{4567}{\K}$. This temperature at the bottom of the simulation domain is then translated to an adiabatic temperature profile for the atmospheric layers below \SI{10}{\bar} and is used to kick off the \texttt{steroids} model.

In order to force the simulation to continue with this new temperature profile for pressure levels below \SI{10}{\bar}, we developed a method that would smoothly force the temperature towards the predicted state during runtime. This method consists of a time-based smoothing, where the temperature is forced towards the final state by splitting the total change of temperature over a period of $\SI{10}{\day}$. We make sure that the temperature is changed by the expected amount, but still modulate the change rates by a sine function $\frac{\mathrm{d}T}{\mathrm{d}t}\propto\sin\left(\pi{\frac{t}{\SI{10}{\day}}}\right)$ to smoothen the transition. A linear vertical smoothing between \SI{1}{\bar} and \SI{10}{\bar} ensures that the upper atmosphere above \SI{1}{\bar} stays untouched from the artificial forcing, which should only change the temperature in the deep layers. These measures are only used for the period of \SI{10}{\day}, and their only purpose is to force the temperature of the \texttt{steroids} model towards its new initial condition. The temperature forcing after \SI{40010}{\day} is then given by standard radiative heating and cooling as computed self-consistently within the GCM.

\FloatBarrier
\section{Models of WASP-76b}\label{sec:atmo}
\begin{figure}
        \includegraphics[width=.45\textwidth]{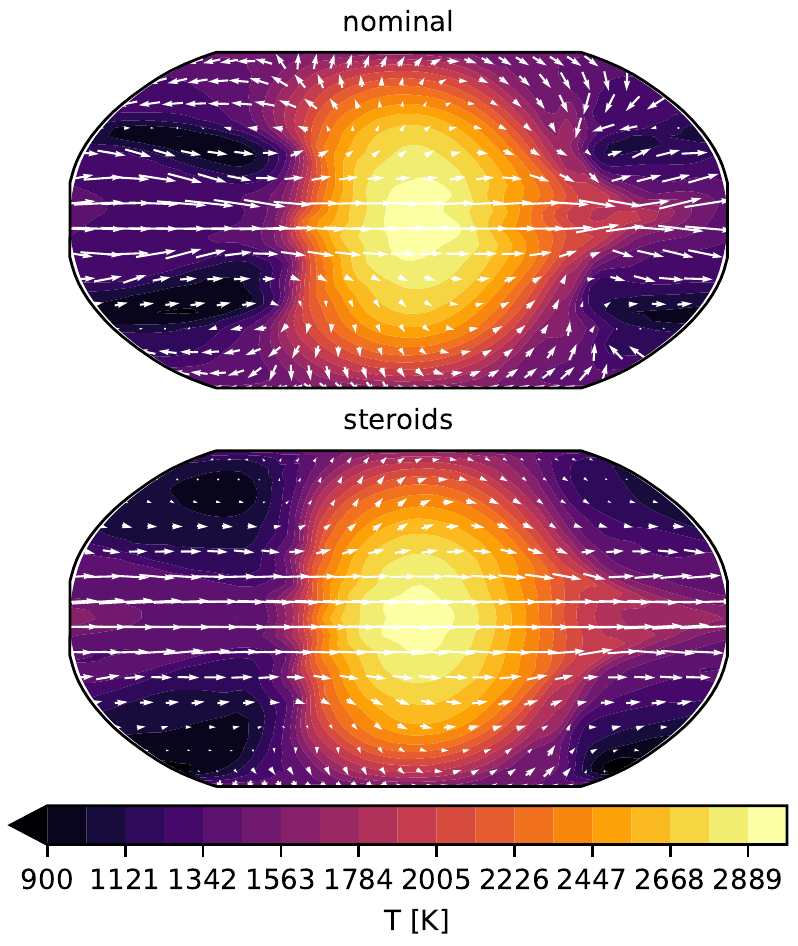}
        \caption{Final isobaric temperature-slices at $p\approx\SI{1.2}{\bar}$ for the \texttt{nominal} and \texttt{steroids} model. The temperature has been averaged over the last \SI{100}{\day}. The substellar point is located in the center of the plot.}
        \label{fig:T}
\end{figure}
\begin{figure*}
        \includegraphics{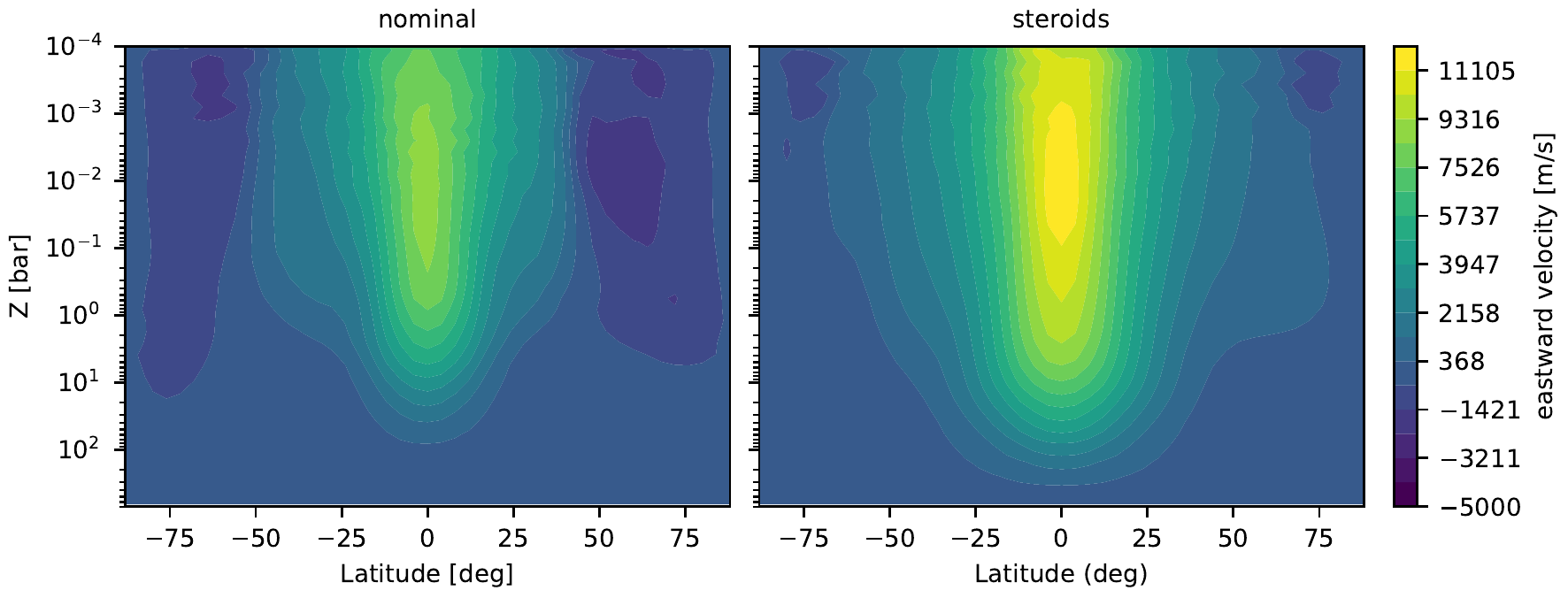}
        \caption{Zonal mean eastward velocity for the \texttt{nominal} and the \texttt{steroids} model.}
        \label{fig:U}
\end{figure*}
\begin{figure*}
        \centering
        \includegraphics{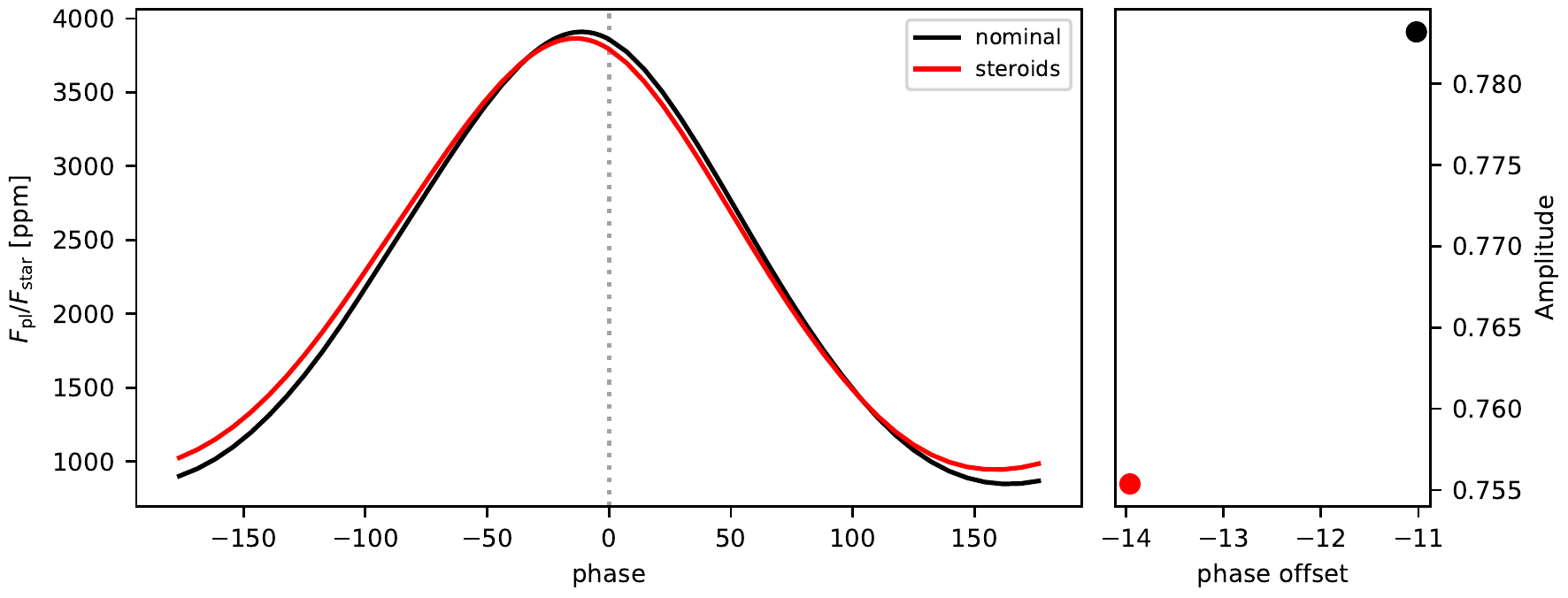}
        \caption{Phase curve and phase offset. Left: Phase curve at $\lambda=\SI{4.5}{\mu\m }$, computed using \texttt{prt\_phasecurve}. The dashed gray line corresponds to the center of the dayside. Right: Difference between the dayside center and the maximum of the phase curve.}
        \label{fig:phasecurve}
\end{figure*}

We found in \citet{Schneider2022GCM} that the influence of the deep atmospheric state on the temperature profile in the upper atmospheric layers is very limited. As we see in Fig.~\ref{fig:timeevol}, the global average of the temperature in the \texttt{steroids} and \texttt{nominal} model only diverges for pressures above \SI{1}{\bar}. We show an isobaric map of the temperature at \SI{1}{\bar} in Fig.~\ref{fig:T}, where we can see that the \texttt{steroid} and \texttt{nominal} model look similar, yet not identical. This becomes especially apparent in the plots of the zonal mean of the zonal wind velocity in Fig.~\ref{fig:U}. The winds in the \texttt{steroids} model are much stronger than the winds in the \texttt{nominal} model. This will have an observable impact, resulting in higher hot spot offsets  and greatly impacting high-resolution Doppler observations.

These strong differences in wind speed were not observed in the simulations of \citet{Schneider2022GCM}, where we did not compare models with such differences in temperature. The transition to higher wind speeds might be caused by the increase in the weather layer, which is caused by larger (relative) horizontal anisotropies in the deep atmosphere, which affects the jet strength \citep{Mayne2017GCM}. However, we note that the kinetic energy makes up less than 1\%\ of the energy budget in our models. It is therefore also possible that the reason for the differences is simply rooted in the particularly poor conservation of angular momentum in \texttt{MITgcm} \citep{Polichtchouk2014GCM}. Further study is needed to entangle the dynamical differences in models with cold and hot interiors. 

To quantify the hot spot offset, we computed phase curves using \texttt{prt\_phasecurve}\footnote{available at: \url{https://prt-phasecurve.readthedocs.io/en/latest/}}. \texttt{prt\_phasecurve} computes 288 intensity fields at the top of the atmosphere using \texttt{petitRADTRANS} \citep{Molliere20191Dmodel,Molliere20201Dmodel} on a longitude latitude grid with $15^\circ$ resolution and for 20 angles instead of the nominal 3 angles that \texttt{petitRADTRANS} uses. These individual spectra are then integrated by taking the individual angles into account (i.e., without the assumption of isotropic irradiation). 

The difference in the offset shift can be seen in the thermal phase curve in Fig.~\ref{fig:phasecurve}. The overall shape of the phase curves in the \texttt{nominal} and \texttt{steroids} model is very similar. However, we can see that the day--night contrast is higher in the case of the \texttt{nominal} model, which can be explained by the strong winds in the \texttt{steroids} model that are more efficient in equilibrating the temperature in the \texttt{steroids} model. Similarly, we can see that the hot spot offset is greater in the \texttt{steroids} model, which is another outcome of the strong winds that transport energy further, before it can be reradiated.

\end{appendix}
\end{document}